\documentclass[%
reprint,
 amsmath,amssymb,
 aps,
prb,
]{revtex4-1}

\usepackage{graphicx}
\usepackage{dcolumn}
\usepackage{bm}
\usepackage{tabularx}
\usepackage{array,booktabs}
\usepackage{hyperref}
\usepackage{color}
\usepackage{booktabs}
\def\diag{\mathop{\rm diag}\nolimits}
\def\1{\mbox{1}\hspace{-0.25em}\mbox{l}}

\begin{document}

\preprint{APS/123-QED}

\title{Giant surface Edelstein effect in $d$-wave superconductors}

\author{Yuhei Ikeda$^1$}
\email{ikeda.yuhei.32w@st.kyoto-u.ac.jp}
\author{Youichi Yanase$^{1,2}$}%
\email{yanase@scphys.kyoto-u.ac.jp}
 
\affiliation{%
 $^1$Department of Physics, Graduate School of Science, Kyoto University, Kyoto 606-8502, Japan\\
 $^2$Institute for Molecular Science, Okazaki 444-8585, Japan 
}%

\date{\today}

\begin{abstract}
Edelstein effect is useful for electric control of magnetic moment. 
However, Joule heating created by a dissipative current is harmful for its practical applications. In this paper, we investigate two-dimensional noncentrosymmetric superconductors (NCSs) with either $s$-wave or $d$-wave symmetry, and demonstrate that a surface Edelstein effect is significantly enhanced in $d$-wave NCSs. The origin of the enhancement is attributed to surface Majorana states characteristic of gapless spin-singlet superconductors. 
In the view of superconducting spintronics, this result would give a route to magnetic domain switching by dissipationless supercurrent. 
We discuss a possible experimental observation in cuprate superconductor heterostructures and heavy fermion superlattices.
\end{abstract}
\maketitle


\section{\label{sec:introduction}Introduction\protect}

Recent studies have uncovered various phenomena arising from spin-orbit coupling, particularly in electron systems lacking inversion symmetry \cite{manchon2015new}.
For instance, superconductivity without inversion symmetry~\cite{bauer2012non}, spin Hall effect \cite{murakami2003dissipationless,sinova2004universal, RevModPhys.87.1213}, magnetoelectric/Edelstein effect \cite{edelstein1990spin}, and topological insulators/superconductors \cite{RevModPhys.82.3045,RevModPhys.83.1057} have been central topics in modern condensed matter physics ranging from superconductivity to spintronics. 

In the research of spintronics, the Edelstein effect has been investigated as a principle for electric control of magnetization.
Spin-orbit torque by the Edelstein effect was proposed \cite{manchon2009theory}, and the magnetic domain switching by applying an electric current was demonstrated in ferromagnet \cite{chernyshov2009evidence,miron2010current,jiang2019efficient} and antiferromagnet \cite{wadley2016electrical}.
However, it is widely known that a large amount of electric current density is needed for magnetic domain switching.
Therefore, Joule heating created by the dissipative electric current is one of main obstacles for spintronics applications.
Thus, it is going to be important to study magnetoelectric/Edelstein effect due to non-dissipative electric current, one of which is supercurrent~\cite{PhysRevLett.75.2004,edelstein2003triplet,yip2002two,yip2005magnetic,fujimoto2007electron,he2019spin,he2020magnetoelectric}.

On the other hand, in the research field of superconductivity, clarification of topological aspects has been intensively conducted for the two decades \cite{sato2016majorana}. 
A remarkable discovery is the concept of topological superconductivity and accompanied Majorana fermion \cite{PhysRevB.61.10267,kitaev2001unpaired}. 
Topologically protected surface states appear not only in the gapped topological superconductors in a usual sense but also in gapless superconductors, where Andereev bound states are characterized by a low-dimensional topological invariant~\cite{yada2011surface,sato2011topology,schnyder2011topological}. 
For instance, noncentrosymmetric $d$-wave superconductors show a unique flat dispersion, which is called surface Majorana flat band \cite{yada2011surface,sato2011topology,schnyder2011topological,daido2017majorana}.
The Majorana flat band gives rise to a singular surface density of states at zero energy.
Therefore, it is expected that a large transport response may be carried by the surface state. 

\begin{figure}
\includegraphics[width=60mm]{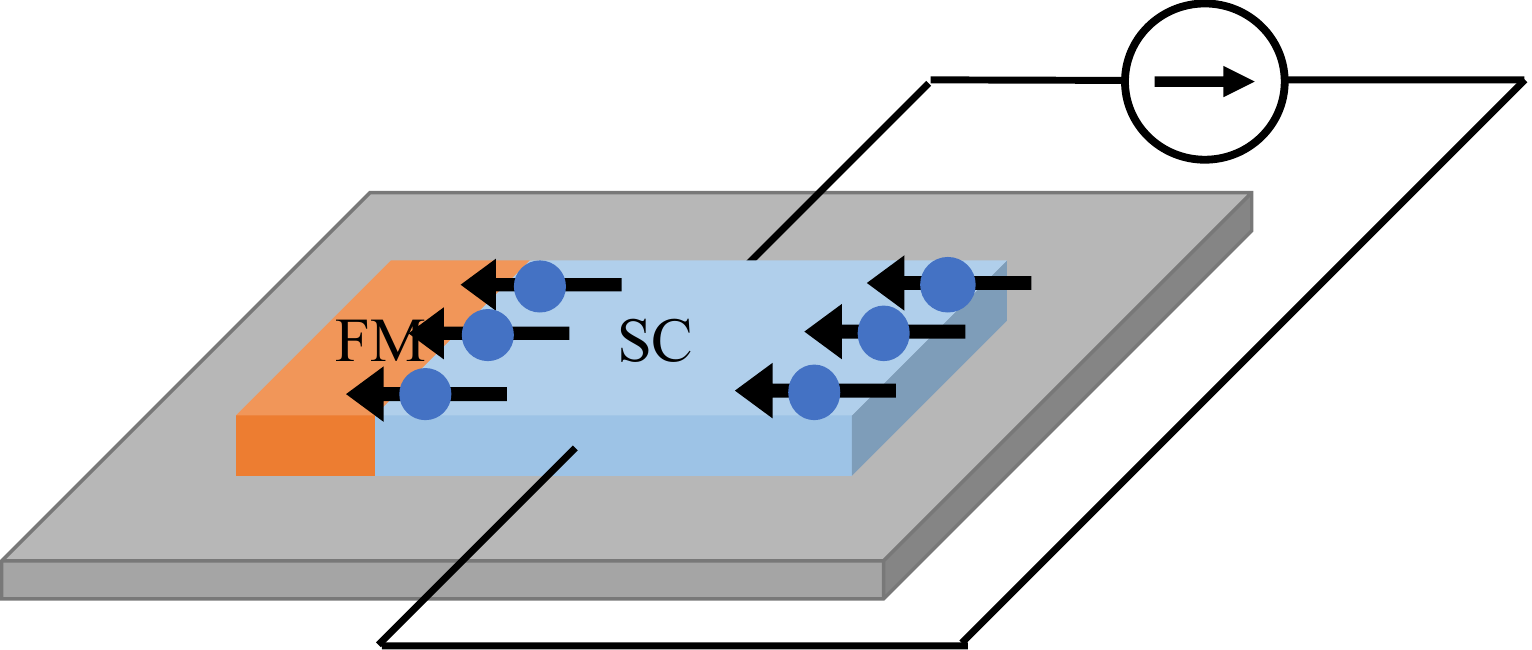}
\caption{\label{fig:SC}(Color online) Illustration of the surface Edelstein effect. 
A $d$-wave superconductor thin film such as cuprate \cite{bollinger2011superconductor} and heavy fermion CeCoIn$_5$ \cite{mizukami2011extremely} hosts spin polarization under an electric supercurrent. The spin polarization is enhanced near edges and its direction is equivalent between the opposite edges. Junction to ferromagnet is illustrated for a proposal of large spin-orbit torque. 
}
\end{figure}

Motivated by the above advances, we study the superconducting Edelstein effect in noncentrosymmetric $d$-wave superconductors.
The superconducting Edelstein effect was previously investigated with considering $s$-wave superconductors~\cite{PhysRevLett.75.2004,edelstein2003triplet,yip2002two,yip2005magnetic,fujimoto2007electron,he2019spin,he2020magnetoelectric}. Distinct from the previous studies, we clarify the surface response mainly caused by the topological surface states. Although the superconducting Edelstein effect is tiny in bulk superconductors~\cite{PhysRevLett.75.2004,yip2002two,yip2005magnetic,fujimoto2007electron,he2019spin}, we show that the response is significantly enhanced near the surface. The response may be relevant for superconducting spintronics research because spin manipulation is performed near the boundary of two different materials~\cite{chernyshov2009evidence,miron2010current,jiang2019efficient} (see illustration in Fig.~\ref{fig:SC}). 

In this paper, we investigate (quasi-)two-dimensional Rashba $d$-wave superconductors. 
The presence of the Rashba spin-orbit coupling in cuprate and heavy fermion $d$-wave superconductors has been clarified by recent experiments \cite{gotlieb2018revealing,shimozawa2016kondo,naritsuka2017emergent}. 
In Sect.~\ref{sec:model}A, we introduce a model of current flowing $d$-wave superconductors with an open boundary.
The superconducting Edelstein effect in this model is formulated in Sect.~\ref{sec:model}B.
In Sect.~\ref{sec:model}C, we revisit the Kubo formula for the Edelstein effect in the normal state for a comparison.
In Sect.~\ref{sec:result}, we show numerical results which reveal significant enhancement of the surface Edelstein effect, compared with bulk Edelstein effects in normal and superconducting states.
In Sect.~\ref{sec:majorana}, the giant surface Edelstein effect is attributed to the surface Majorana states. This view is supported by comparison of $s$-wave and $d$-wave superconducting states. In Sect.~\ref{sec:proposal}, we conclude the paper with discussing an experimental setup and candidate materials.

\section{\label{sec:model}Formulation}

\subsection{Model for noncentrosymmetric $d$-wave SCs}
A setup of the system is schematically shown in Fig.~\ref{fig:SC}. We consider a noncentrosymmetric $d$-wave superconductor under an applied electric current. With an inversion asymmetry, Rashba spin-orbit coupling appears. Then, the superconducting Edelstein effect, in which supercurrent flow induces magnetization, occurs \cite{PhysRevLett.75.2004}. In order to describe this phenomenon, we introduce a Bogoliubov de-Gennes (BdG) Hamiltonian
\begin{align}
 H &=\sum_{\bm{k},s}\xi(\bm{k})c_{\bm{k},s}^\dag c_{\bm{k},s} +\alpha_{\mathrm{R}}\sum_{\bm{k},s,s'} \Big(\bm{g}(\bm{k})\cdot \bm{\sigma}\Big)_{s,s'}c_{\bm{k},s}^\dag c_{\bm{k},s'}
 \notag\\&+ \sum_{\bm{k},s,s'}\Big(\Delta (\bm{k})c_{\bm{k+q},s}^\dag c_{\bm{-k+q},s'}^\dag+\rm{H.c.}\Big).
\end{align}
The first term describes a kinetic energy measured from a chemical potential $\xi_{\bm{k}}=-2t(\cos k_x + \cos k_y) - \mu$, $t$ is a hopping parameter in the tight-binding model, $\mu$ is the chemical potential, $s$ is an index of spin, and the lattice constant $a$ is set to unity. The second term is the Rashba spin-orbit coupling, $\alpha_{\mathrm{R}}$ is the coupling constant, $\bm{g}(\bm{k}) = (-\sin k_y,\sin k_x,0)$, and $\bm{\sigma}$ is the Pauli matrix $\bm{\sigma} = (\sigma_x,\sigma_y,\sigma_z)$.
The last term arises from Cooper pairs having center of mass momentum $2\bm{q}$. In this setup non-dissipative supercurrent is carried by Cooper pairs with a finite momentum. Conversely, when a supercurrent flows, the superconducting order parameter acquires a phase gradient along the current flowing direction. Distinct from previous studies~\cite{PhysRevLett.75.2004,edelstein2003triplet,yip2002two,yip2005magnetic,fujimoto2007electron,he2019spin,he2020magnetoelectric}, in this paper we consider a $d$-wave superconducting order parameter, $\Delta(\bm{k}) =\Delta_0\sin k_x\sin k_y$. As we show later, $d$-wave superconductors cause a giant surface Edelstein effect due to topological surface states although the bulk Edelstein effect is not sensitive to the symmetry of superconductivity. Although we here consider $d_{xy}$-wave superconductivity for simplicity, our analysis reveals that the large surface Edelstein effect universally occurs in $d$-wave superconductors.

To evaluate physical quantities at the surface and bulk in a coherent way, we set an open boundary condition along the (100) direction. In this condition, a wave number $k_x$ is not a good quantum number, and therefore, a Fourier transformation along the $x$ direction is applied to the model. We consider an electric current flowing along the $y$ direction, and the center of mass momentum $\bm{q}$ restricted to the $y$ direction $\bm{q} = q\bm{\hat{y}}$ is adopted.

In this setup the BdG Hamiltonian is spanned by the Nambu spinor
\begin{align}
\Phi_{i_x,k_y} = 
\begin{pmatrix}
c_{k_y+q,i_x,\uparrow}\\
c_{k_y+q,i_x,\downarrow}\\
c_{-k_y+q,i_x,\uparrow}^\dag\\
c_{-k_y+q,i_x,\downarrow}^\dag
\end{pmatrix}, 
\end{align}
where $i_x$ is a spacial coordinate. By using this basis, the Hamiltonian is described as a $4N \times 4N$ matrix (spin space $\bigotimes$ particle-hole space $\bigotimes$ real space)
\begin{equation}
 \mathcal{H}(k_y) = \left(
    \begin{array}{@{\,}cccc@{\,}}
       \mathcal{H}_{1,1}(k_y)   & \mathcal{H}_{1,2}(k_y)  &  \hdots  &  \mathcal{H}_{1,N}(k_y) \\
       \mathcal{H}_{2,1}(k_y)   & \mathcal{H}_{2,2}(k_y)  &  \hdots  &  \mathcal{H}_{2,N}(k_y) \\
        \vdots    & \vdots    &\ddots   &   \vdots   \\
        \mathcal{H}_{N,1}(k_y) &   \mathcal{H}_{N,2}(k_y) & \hdots  &   \mathcal{H}_{N,N}(k_y) \\
    \end{array}
    \right), 
\end{equation}
with
\begin{align}
&\mathcal{H}_{i_x,j_x}(k_y) = \notag \\
&  \begin{pmatrix}
      H_{\mathrm{N}}(k_y+q;i_x,j_x) & \Delta(k_y;i_x,j_x) (i\sigma_y) \\
      \Delta^*(k_y;j_x,i_x) (i\sigma_y)^{\dag} & -H_{\mathrm{N}}(-k_y+q;j_x,i_x)^T\\
    \end{pmatrix}, 
\end{align}
where $H_{\mathrm{N}}$ is the Hamiltonian in the normal state. An explicit form is
\begin{align}
H_{\mathrm{N}}(k_y;i_x,j_x) = &\Big[-2t \big(\frac{\delta_{i_x,j_x+1}+\delta_{i_x+1,j_x}}{2} \notag\\
&+\cos k_y \delta_{i_x,j_x}\big) - \mu \delta_{i_x,j_x}\Big]\1_{2\times2}\\
&+ \notag \alpha_{\mathrm{R}} \bm{g}(k_y+q;i_x,j_x)\cdot \bm{\sigma}, 
\end{align}
where $\delta_{i_x,j_x}$ is the Kronecker's delta.
In this model, the g-vector has the form
\begin{equation}
\bm{g}(k_y;i_x,j_x) =  \Big(-\sin k_y \delta_{i_x,j_y} ,\frac{\delta_{i_x,j_x+1}-\delta_{i_x+1,j_x}}{2i}, 0\Big).
\end{equation}
The Fourier transformed superconducting order parameter is obtained as
\begin{align}
\Delta(k_y;i_x,j_x) = |\Delta_0|\Big(\frac{\delta_{i_x,j_x+1}-\delta_{i_x+1,j_x}}{2i}\Big)\sin k_y\1_{2\times 2}.
\end{align}
The Hamiltonian is described as
\begin{equation}
    H = \frac{1}{2}\sum_{i_x,j_x,k_y}\Phi^\dag_{i_x,k_y}\mathcal{H}_{i_x,j_x}(k_y)\Phi_{j_x,k_y}.
\end{equation}

\subsection{Superconducting Edelstein effect}

In noncentrosymmetric metals and superconductors an electric current induces spin magnetization~\cite{edelstein1990spin,PhysRevLett.75.2004}. This phenomenon is called Edelstein effect and it is described as $M_\mu = - \gamma_{\mu\nu} J_{\nu}$. In Rashba systems polar inversion symmetry breaking leads to a finite $\gamma_{xy}$ and the Edelstein effect is 
\begin{equation}\label{superEdelstein}
    M_x = -\gamma_{xy} \, J_y. 
\end{equation}
The magnitude of the Edelstein effect is characterized by the Edelstein coefficient $\gamma_{xy}$. For more general cases, a group theoretical analysis of the response tensor $\gamma_{\mu\nu}$ was conducted \cite{he2020magnetoelectric} and corresponding multipole moment has been clarified \cite{PhysRevB.98.245129}.

To evaluate the Edelstein effect we calculate the spin magnetization and electric current. They are obtained as statistical average of their quantum mechanical operators, which are space dependent as 
\begin{align}
    S_x(i_x) &= \sum_{k_y}\Phi_{i_x,k_y}^{\dag}s_x \Phi_{i_x,k_y},\\
    J_y(i_x) &= \sum_{k_y}\Phi_{i_x,k_y}^{\dag}j_y(k_y) \Phi_{i_x,k_y}. 
\end{align}
Here we defined 
\begin{align}
    s_x &= \frac{1}{2}
    \begin{pmatrix}
    \sigma_x    &   O   \\
    O           &   -\sigma_x^T \\
    \end{pmatrix},\\
    j_y(k_y) &= \frac{e}{2}
    \begin{pmatrix}
    \left.\frac{\partial H_{\mathrm{N}}}{\partial k_y}\right|_{k_y+q} &   O   \\
        O                                   &   -\Big(\left.\frac{\partial H_{\mathrm{N}}}{\partial k_y}\right|_{-k_y+q}\Big)^{T}\\
    \end{pmatrix}.
\end{align}
These operators are obtained by projecting the spin and current operators to the spacial coordinate $i_x$. 

To calculate the statistical average we diagonalize the Hamiltonian by using a $4N\times4N$ unitary matrix
\begin{equation}
 U(k_y) = \left(
    \begin{array}{@{\,}cccc@{\,}}
        U_{1,1}(k_y)   & U_{1,2}(k_y)  &  \hdots  &   U_{1,N}(k_y) \\
        U_{2,1}(k_y)   & U_{2,2}(k_y)  &  \hdots  &   U_{2,N}(k_y) \\
        \vdots    & \vdots    &\ddots   &   \vdots   \\
        U_{N,1}(k_y) &   U_{N,2}(k_y) &   \hdots  &   U_{N,N}(k_y) \\
    \end{array}
    \right), 
\end{equation}
where $U_{l,m}$ are $4\times4$ matrices and $N$ is the length of the system along the $x$ axis. After the unitary transformation the Hamiltonian which we consider in this paper turned into a diagonal matrix
\begin{equation}
H = \frac{1}{2}\sum_{i_x,j_x,m,k_y}\Phi^\dag_{i_x,k_y}U^\dag_{i_x,m}(k_y)H^{(d)}_{m}(k_y)U_{m,j_x}(k_y)\Phi_{j_x,k_y}.
\end{equation}
Here, $H^{(d)}_{m}(k_y)=\diag{(\varepsilon_{1,m}(k_y),\varepsilon_{2,m}(k_y),\varepsilon_{3,m}(k_y),\varepsilon_{4,m}(k_y))}$, and $\varepsilon_{n,m}(k_y)~ (1 \le n \le 4)$ are eigen energies of Bogoliubov quasiparticles.
We define operators 
\begin{align}
\sum_{i_x}U_{m,i_x}(k_y)\Phi_{i_x,k_y}=
\begin{pmatrix}\alpha_{k_y+q,m,\uparrow}\\
\alpha_{k_y+q,m,\downarrow}\\
\alpha_{-k_y+q,m,\uparrow}^\dag\\
\alpha_{-k_y+q,m,\downarrow}^\dag
\end{pmatrix},
\end{align}
for the Bogoliubov quasiparticles. We obtain expectation values for spin and electric current as
\begin{align}
    \langle S_x(i_x)\rangle &= \sum_{k_y,n,m}f(\varepsilon_{n,m}(k_y))\big[U_{m,i_x}(k_y)s_x U^{\dag}_{i_x,m}(k_y)\big]_{n,n},\\
    \langle J_y(i_x)\rangle &= \sum_{k_y,n,m}f(\varepsilon_{n,m}(k_y))\big[U_{m,i_x}(k_y)j_y(k_y)U^{\dag}_{i_x,m}(k_y)\big]_{n,n}, 
\end{align}
with the Fermi distribution function $f(x)$. 

Since we are interested in the spacial dependence of magnetization, we characterize the Edelstein effect in a space-dependent form, 
\begin{equation}
    \gamma_{xy}(i_x) = -\mu_{\mathrm B} \langle S_x(i_x) \rangle / J_y, 
\end{equation}
with assuming the $g$-factor $g=2$.
Because most of electric current flows in the bulk, the averaged electric current density is approximated as 
\begin{equation}
    J_y = \sum_{i_x}\langle J_y(i_x) \rangle/N \simeq \langle J_y(N/2)\rangle.
\end{equation}

In the next section, we compare the surface Edelstein effect and the bulk Edelstein effect, which are characterized by $\gamma_{xy}(i_x)$ for $i_x = 1$ and $i_x = N/2$, respectively. 

\subsection{Edelstein effect in normal state}

 To compare the superconducting Edelstein effect with that in the normal state, we also calculate a magnitude of the Edelstein effect in the normal state. There is no reason which ensures similar magnitudes between the superconducting state and the normal state because the source field is essentially different. 
In the normal state magnetization is induced by an electric current with dissipation, while the superconducting Edelstein effect is owing to the dissipationless supercurrent. 

To evaluate the response with dissipation we here use Kubo formula. 
In the linear response theory~\cite{bauer2012non}, the Edelstein effect is described as
\begin{equation}\label{chi_xy}
    M_x = -\chi_{xy}E_y, 
\end{equation}
where $E_y$ is an electric field along the $y$ axis. The coefficient $\chi_{xy}$ is calculated by using the standard Kubo formula, 
\begin{equation}
    \chi_{xy}(\omega) =  \frac{1}{i\omega}\left.K_{xy}(i\omega_n)\right|_{i\omega_n \rightarrow \omega + i \delta}.
\end{equation}
The response function $K_{xy}$ is given by
\begin{equation}
    K_{xy}(i\omega_n) = \frac{g\mu_{\mathrm{B}} }{2}\int_0^{1/T}d\tau  \langle T_\tau S_x(\tau)J_y(0)\rangle e^{i\omega_n\tau}, 
\end{equation}
where $g$ is the Lande's $g$ factor. Here, spin and current operators are defined as
\begin{align}
    S_x &= \sum_{\bm{k}} C_{\bm{k}}^\dag\sigma_xC_{\bm{k}}, \\
    J_y &= e\sum_{\bm{k}} C_{\bm{k}}^\dag v_y(\bm{k})C_{\bm{k}},
\end{align}
with $C_{\bm{k}}=(c_{\bm{k},\uparrow},c_{\bm{k},\downarrow})^T$ and $v_y(\bm{k})=\partial_{k_y}H_{\mathrm{N}}(\bm{k})$. As a result of a straightforward calculation we obtain the coefficient of Edelstein effect in the Rashba model,
\begin{equation}
    \chi_{xy} = \frac{ge\mu_{\mathrm{B}}}{2}\sum_{\bm{k}}\sum_{\sigma=\pm}\frac{\sin{k_y}}{\sqrt{\sin{k_x}^2+\sin{k_y}^2}}[\sigma v_y^{(\sigma)}(\bm{k})K_{\sigma}(\bm{k})], 
\end{equation}
with
\begin{equation}
    K_{\pm}(\bm{k}) = \frac{1}{\pi}\int_{-\infty}^{\infty} d\epsilon (-f'(\epsilon))[\text{Im}\mathcal{G}_{\pm}^R(\bm{k},\epsilon)]^2. 
\end{equation}
Here, $\mathcal{G}^R_\pm(\bm{k},\epsilon) = [\epsilon - E_\pm(\bm{k}) +i/2\tau_\pm(\bm{k})]^{-1}$ is a retarded Green function for quasi-particles with energy, $E_\pm(\bm{k}) = \xi(\bm{k}) \pm \alpha_{\mathrm{R}}(\sin^2 k_x + \sin^2 k_y)^{1/2}$, and velocity, $v_y^{(\pm)}(\bm{k})= \partial _{k_y}E_{\pm}(\bm{k})$.
In the following calculations, we assume that the damping constant is not dependent on spin, $\tau_\pm(\bm{k})=\tau$, for simplicity. 
Then the integration turns into
\begin{equation}
     K_{\pm}(\bm{k}) = \tau[-f'(E_\pm(\bm{k}))], 
\end{equation}
and we obtain
\begin{align}
     \chi_{xy} = e\mu_{\mathrm{B}}\tau \sum_{\bm{k}}\sum_{\sigma=\pm} & \frac{\sin{k_y}}{\sqrt{\sin{k_x}^2+\sin{k_y}^2}} \notag \\
    & \times \sigma v_y^{(\sigma)}(\bm{k})[-f'(E_{\sigma}(\bm{k}))], 
\end{align}
by adopting $g=2$.

In order to rewrite Eq.~\eqref{chi_xy} to the form of the Edelstein effect Eq.~\eqref{superEdelstein} we also calculate the electric conductivity in the same manner. The result is 
\begin{equation}
    \sigma_{y} = e^2 \tau \sum_{\bm{k}}\sum_{\sigma=\pm} v^{(\sigma)}_y(\bm{k})^2[-f'(E_\sigma(\bm{k}))].
\end{equation} 
The Ohm's law $J_y = \sigma_y E_y$ leads to 
\begin{equation}\label{normal}
    M_x = -\frac{\chi_{xy}}{\sigma_y}\sigma_y E_y = -\gamma_{xy} J_y. 
\end{equation}
Thus, we obtain the Edelstein coeffiient $\gamma_{xy}$ in the normal state as
\begin{equation}\label{normalgamma}
\gamma_{xy} = \frac{\chi_{xy}}{\sigma_y}.
\end{equation}
Note that $\gamma_{xy}$ is an intrinsic quantity in the sense that it does not depend on the quasiparticle's life time $\tau$ although the response comes from a dissipative process. Therefore, we can compare the magnitudes of Edelstein effect in the superconducting and normal states without relying on an extrinsic parameter $\tau$.

\section{\label{sec:result}Numerical results}
In this section we show the numerical results revealing significant enhancement of the Edelstein effect in $d$-wave superconductors at surfaces. 

\subsection{Giant surface Edelstein effect}

\begin{figure*}[htbp]
\includegraphics[width=140mm]{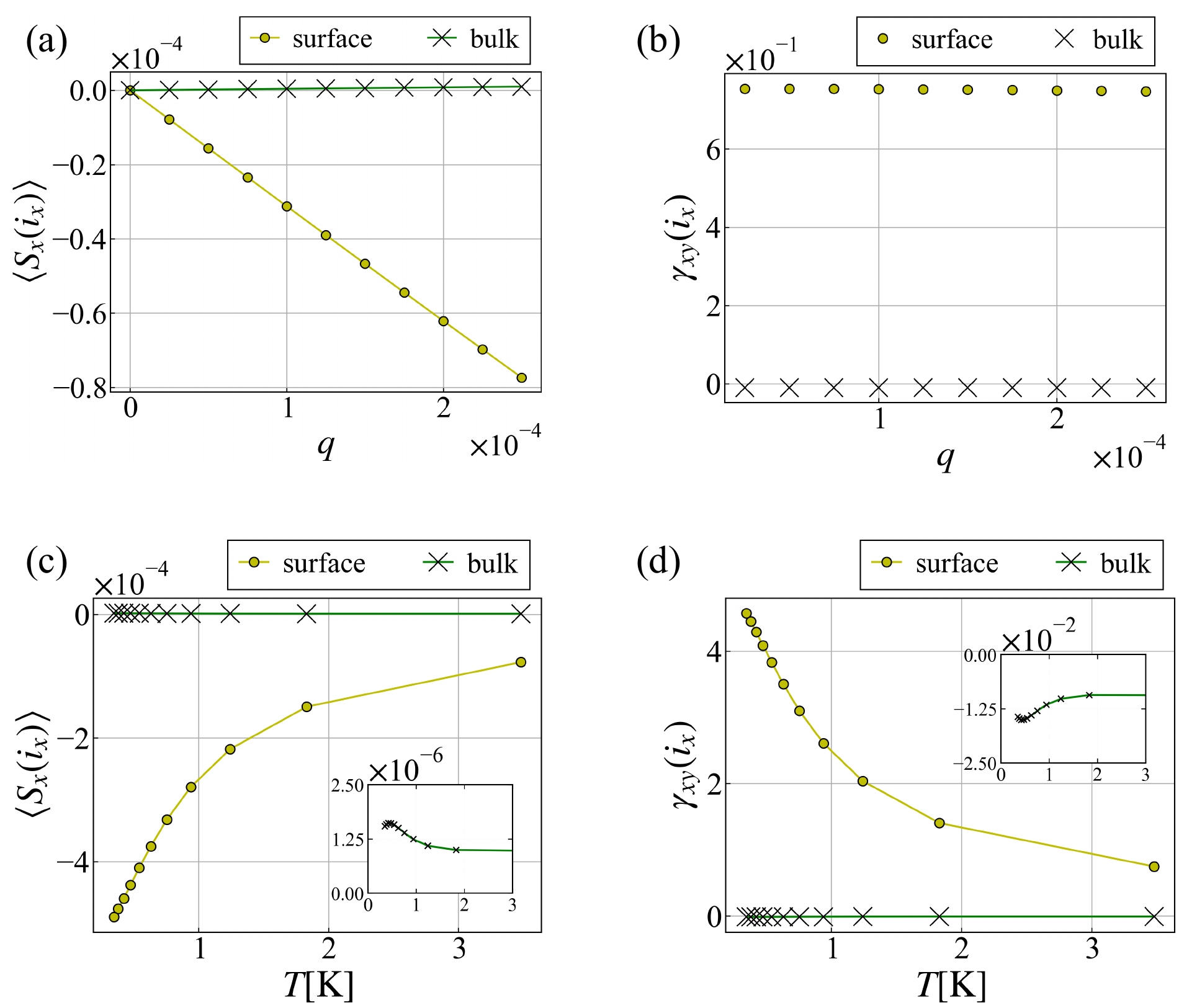}
\caption{(Color online) Comparison of superconducting Edelstein effects between the surface ($i_x=1$, circles with yellow lines) and bulk ($i_x=N/2$, crosses with green lines). Center of mass momentum $q$ dependence of (a) spin expectation value $\langle S_x(i_x) \rangle$ and (b) Edelstein coefficient $\gamma_{xy}(i_x)$. Parameters are $t=1$, $\Delta_0$ = 0.1, $\alpha_{\mathrm{R}}$ = 0.05, $\mu=-2.5$, and $\beta = 1/T = 1000$. We set $e=\mu_{\mathrm{B}}=1$ for simplicity. (c) and (d) show temperature dependence, where the temperatures are estimated by assuming $t=300$ [meV]. In (c) and (d) we adopt $q = 2.5 \times 10^{-4}$. Inset of (c) and (d) are enlarged figures for the bulk magnetization and bulk Edelstein coefficient, respectively.
}
\label{fig:q_dept}
\end{figure*}

First we show the results for the superconducting Edelstein effect. In the following part we adopt the unit of energy $t$ = 1, and set parameters $\Delta_0$ = 0.1, $\alpha_{\mathrm{R}}$ = 0.05, $\mu=-2.5$ unless we mention explicitly. Figure~\ref{fig:q_dept}(a) shows the $q$-dependence of spin expectation value at the surface ($i_x=1$) and bulk ($i_x=N/2$). We see that the spin magnetization at the surface is much larger than that at the bulk. 
Indeed, the magnetization in the bulk region is negligible compared with the surface magnetization. 
Both spin and electric current obey almost linear $q$-dependence in the adopted parameter region, and therefore, the Edelstein coefficient $\gamma_{xy}$ plotted in Fig.~\ref{fig:q_dept}(b) is almost $q$-independent. 
The results in Figs.~\ref{fig:q_dept}(a) and \ref{fig:q_dept}(b) reveal that the Edelstein effect is significantly enhanced at surfaces of noncentrosymmetric $d$-wave superconductors. For the parameters in Fig.~\ref{fig:q_dept}(b) the enhancement factor is more than $80$. As we discuss in the next section the enhancement is attributed to the Majorana flat band which is a topologically protected surface state of $d$-wave superconductors~\cite{yada2011surface,sato2011topology}. 

We also calculated temperature dependences of the magnetization and Edelstein coefficient, which are plotted in Figs.~\ref{fig:q_dept}(c) and \ref{fig:q_dept}(d), respectively. The results indicate further increase of the surface Edelstein effect by decreasing the temperature. In contrast, the bulk Edelstein effect is almost temperature-independent (see inset of Fig.~\ref{fig:q_dept}(c)), when we neglect the temperature dependence of the order parameter. Note that the sign of spin magnetization is opposite between the surface and bulk. 

\begin{figure}[htbp]
\includegraphics[width=70mm]{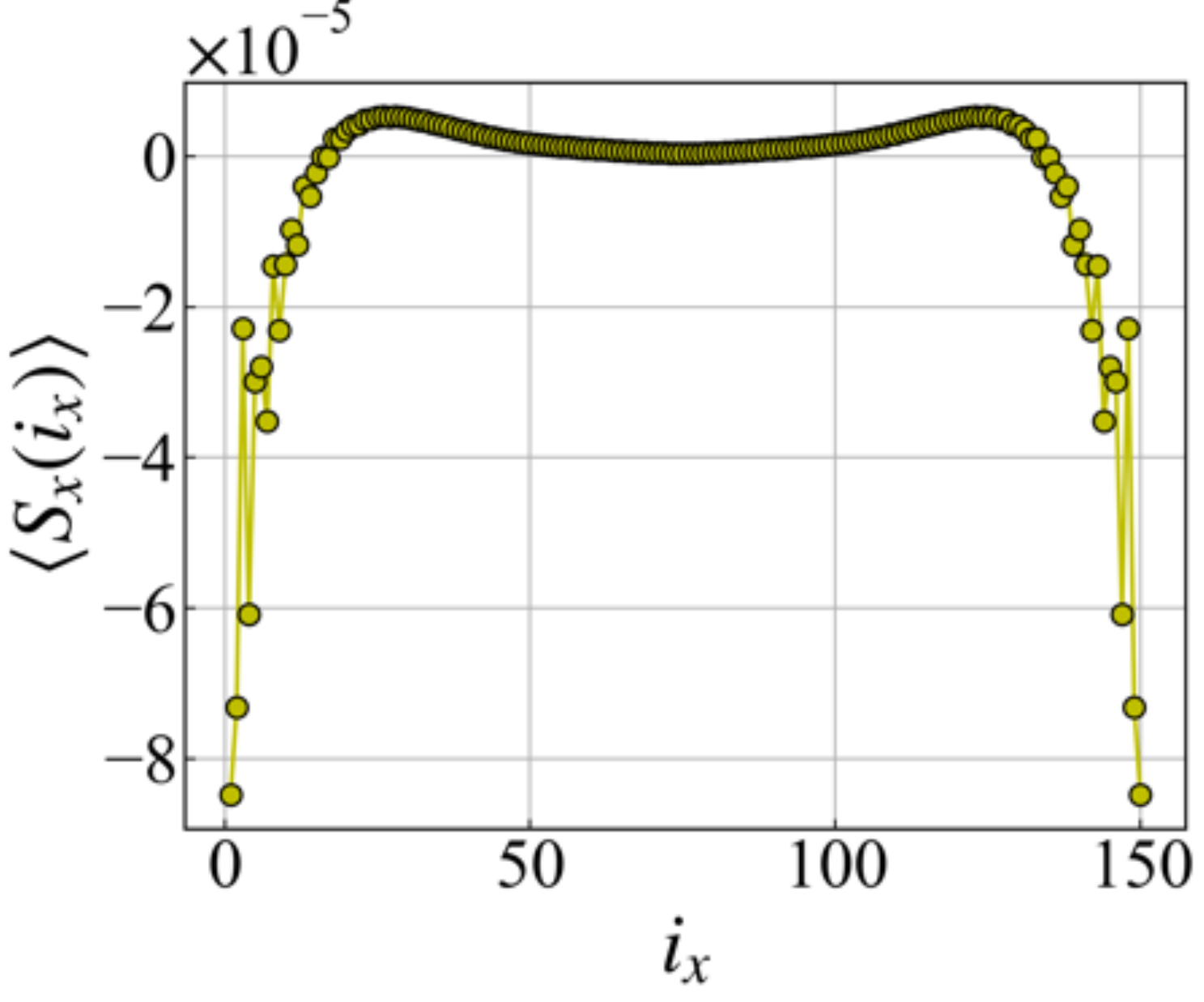}
\caption{(Color online) Site dependence of spin magnetization $\langle S_x(i_x)\rangle$. Parameters are the same as Fig.~\ref{fig:q_dept}(c) at $T = 3.5$ [K].}
\label{fig:site_dept}
\end{figure}

To demonstrate that large magnetization is localized at surfaces, we show the spacial dependence in Fig.~\ref{fig:site_dept}. The magnetization is indeed enhanced around the surfaces and the sign of the magnetization is equivalent between the two surfaces along opposite directions.

\subsection{Quantitative estimation and comparison with normal Edelstein effect}


For a quantitative estimation we adopt a hopping energy $t = 300$ [meV] consistent with cuprate superconductors \cite{PhysRevB.85.064501}. Then, an electric current $J_y = 1$ [A $\cdot$ cm$^{-2}$] induces magnetization $|M_x| =  1 \times 10^{-11} \mu_{\mathrm B}$ in the bulk region. The magnetization at surfaces is typically $100$ times larger than this value.

Next we compare the superconducting Edelstein effect with the normal Edelstein effect. Results of numerical calculations for Eq.~(\ref{normalgamma}) 
show that the normal Edelstein coefficient $\gamma_{xy}$ is almost independent of temperatures, and the magnitude is $\gamma_{xy}= -9 \times 10^{-3}$, slightly smaller than the superconducting Edelstein effect in the bulk region. 
Thus, the surface Edelstein effect in the superconducting state is approximately 100 times larger than the normal Edelstein effect which has been studied in the spintronics research~\cite{manchon2015new,chernyshov2009evidence,miron2010current,jiang2019efficient}. 



Finally we compare our results with a previous theoretical study. A comparison of superconducting and normal Edelstein effects was recently conducted by He \textit{et al}.~\cite{he2019spin}. They have shown that the Edelstein effect is suppressed by the superconducting transition, in contrast to our case. The discrepancy partly comes from the symmetry of superconductivity. Although we are investigating $d$-wave superconductivity, He \textit{et al}. studied conventional $s$-wave superconductivity. In the next section we actually show that the Edelstein effect in the $s$-wave superconducting state is an order of magnitude smaller than the $d$-wave state. Therefore, our result for $s$-wave superconductivity is qualitatively consistent with Ref.~\cite{he2019spin}. However, we note that the definition of the normal Edelstein effect is different between Ref.~\cite{he2019spin} and us. 
In the former, the normal Edelstein effect is formulated by taking the limit $\Delta / \alpha_{\mathrm R}^2 \rightarrow 0$ for the superconducting Edelstein effect. Such formula does not appropriately deal with a dissipative current. In this study, we evaluated the normal Edelstein effect by a standard Kubo formula in which a dissipative current can be taken into account. 


\section{\label{sec:majorana}Role of surface Majorana states}
\begin{figure*}[htbp]
\includegraphics[width=140mm]{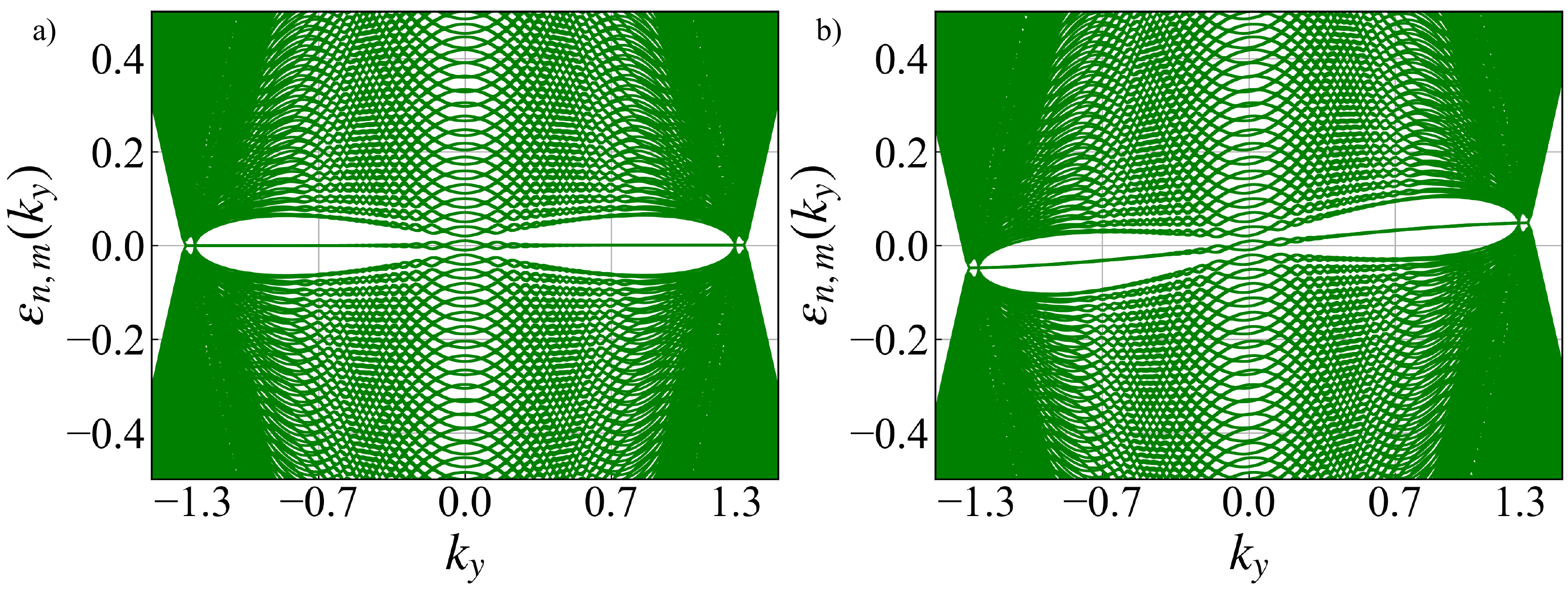}
\caption{\label{fig:band}(Color online) Bogoliubov quasiparticle spectrum for (a) zero and (b) finite Cooper pairs' momentum $q$. (a) shows the superconducting state at rest while (b) shows a supercurrent flowing state. The Majorana flat band in (a) turns into a uni-directional Majorana state in (b). 
We assume a large value of $q$ = 0.02 in (b) for visibility of the uni-directional state.}
\end{figure*}

To discuss the mechanism of the giant surface Edelstein effect, we show the quasiparticle's band structure in the open boundary condition. The energy spectrum in a static state and in a current flowing state are shown in Fig.~\ref{fig:band}(a) and Fig.~\ref{fig:band}(b), respectively. 

In the $d$-wave superconducting state at rest, Andreev bound states form a flat band, when we choose an appropriate open boundary. Importantly, in noncentrosymmetric systems, the Andreev bound states do not have spin degeneracy in a part of the flat band~\cite{sato2011topology}, and it is called Majorana flat band. The Majorana flat band is indeed an origin of the giant surface Edelstein effect, as we discuss below. 

The Majorana flat band actually appears at zero energy in Fig.~\ref{fig:band}(a). We confirmed that the wave functions of the flat band are localized on one of the edges.
The surface states have a topological nature characterized by a one-dimensional winding number~\cite{sato2011topology} 
\begin{equation}
    W(k_y) = - \int^{\pi}_{-\pi} \frac{d k_x}{4\pi i} \mathrm{Tr} \big[\Gamma H_{\mathrm BdG}(\bm{k})^{-1} \partial_{k_x} H_{\mathrm BdG}(\bm{k}) \big], 
\end{equation}
where $\Gamma$ is a chiral operator. The number of surface states at $k_y$ is equivalent to $|W(k_y)|$ ensured by the index theorem. 
On the other hand, when we assume a finite Cooper pairs' momentum $q$, the band structure changes to Fig.~\ref{fig:band}(b). We see that the Majorana flat band state changes to the uni-directional surface state, which is characteristic of gapless topological states with broken time-reversal symmetry. The electric current breaks the time-reversal symmetry in this case.

To identify the origin of the giant surface Edelstein effect, we calculate spin expectation values of the surface states 
\begin{equation}
\langle S_x^{\,\mathrm{edge}}(k_y) \rangle = \sum_{n \in{\mathrm{\{edge\}}}} 
\langle n,k_y | s_x |n,k_y \rangle.
\end{equation}
Here, $|n,k_y \rangle$ denote wave functions of Bogoliubov quasiparticles, and summation for $n$ is restricted to the surface states localized at an edge. 
The spin expectation values of the surface states at four different $k_y$ points ($k_y =-1.3,~-0.7,~0.7,~1.3$) are shown in Table~\ref{tab:edge}. The momentum are chosen so as to satisfy $0.7 < k_1 < 1.3 < k_2$, where $\pm k_1$, $\pm k_2$, and $0$ are wave numbers of nodal points in the superconducting gap. The winding number
which is associated with the degeneracy of surface states changes at these gapless points.
\renewcommand{\arraystretch}{1.2}
\begin{table}
\setlength{\tabcolsep}{8.0pt}
    \centering
    \caption{\label{tab:edge}Spin expectation values of the surface states $\langle S_x^{\mathrm{edge}}(k_y)\rangle$ for $q=0.02$ and the topological winding number $W(k_y)$ for $q=0$ \cite{yada2011surface}. 
    Degeneracy of the surface states is also shown.
}
    \begin{tabular}{l|cccc} \toprule[0.3mm]
    $k_y$         &-1.3       & -0.7   &0.7        &1.3    \\ \midrule[0.1mm]
    $\langle S_x^{\mathrm{edge}}(k_y)\rangle$       &-0.96        &   0.01      &-0.01            &0.96   \\
     $W(k_y)$   &       -1       & -2       &2          &1       \\
     Degeneracy  &       1       &  2       &2          &1       \\
     \bottomrule[0.3mm]
     \end{tabular}
\end{table}
\renewcommand{\arraystretch}{1.0}
Table~\ref{tab:edge} reveals that the spin expectation values are strongly correlated to the topological winding number $W(k_y)$. When the winding number is odd, the number of Majorana fermions is odd so that the spin expectation values must be finite. On the other hand, when the winding number is even, the spin  expectation values of even Majorana fermions can be canceled out. We actually obtain large spin expectation values for $k_1 < |k_y| < k_2$, where $W(k_y)=\pm 1$, although they are negligibly small when $W(k_y)=\pm 2$.

When an electric current is flowing, the flat band gets a finite dispersion, and a uni-directional state appears [Fig.~\ref{fig:band}(b)]. 
Then, the occupation number of quasi-particles is asymmetric with respect to $k_y$. 
For $k_y < 0$ Majorana states with large negative spin expectation values are occupied, while for $k_y > 0$ those with large positive spin expectation values are unoccupied. Therefore, their contributions to the total magnetization are not cancelled, and the flat band with diverging DOS gives rise to the giant surface Edelstein effect.  

\begin{figure}[htbp]
\includegraphics[width=70mm]{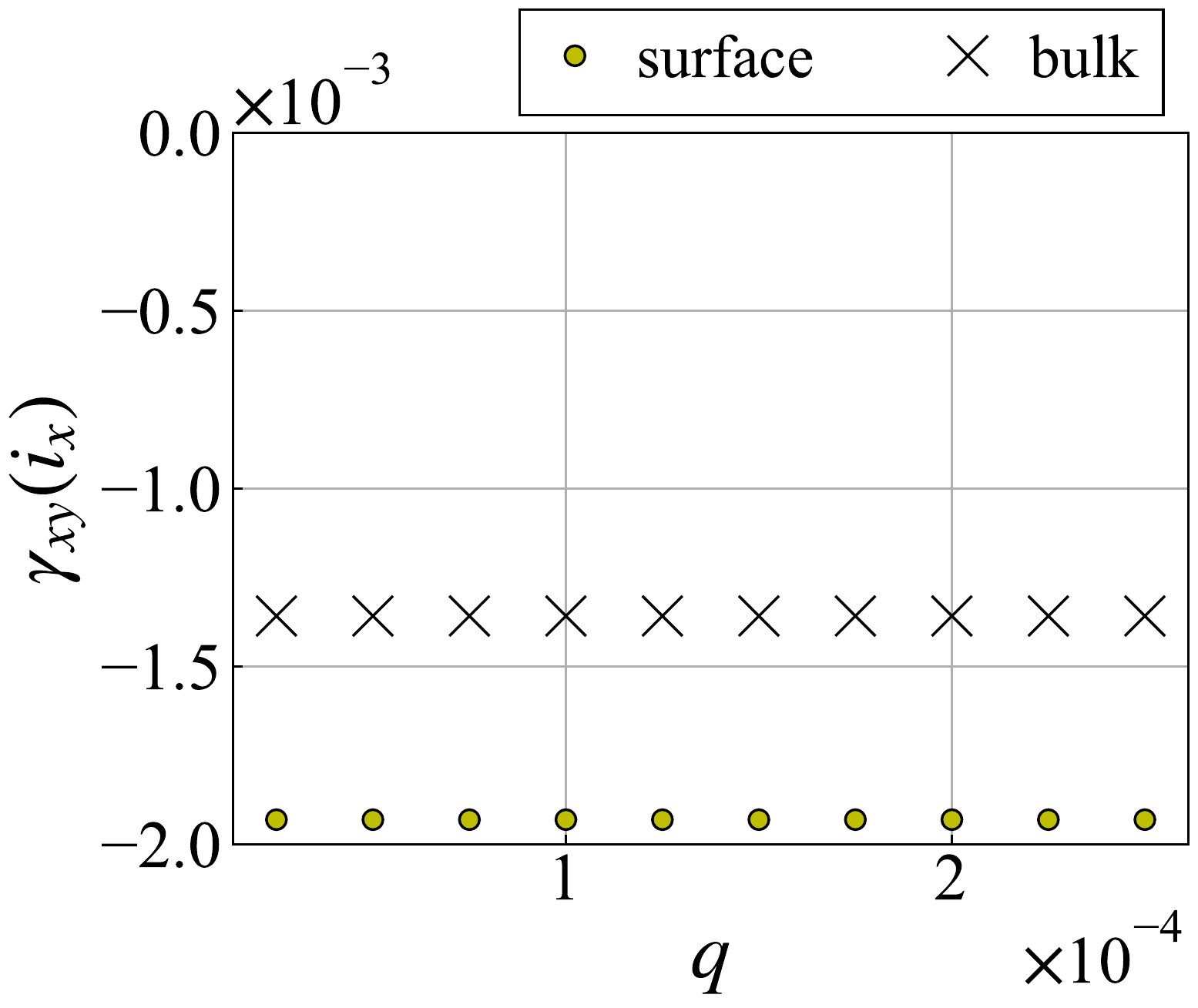}
\caption{(Color online) Edelstein coefficient in a s-wave superconducting state at the surface ($i_x=1$, circles) and bulk ($i_x=N/2$, crosses). 
}
\label{fig:s-wave}
\end{figure}

In order to demonstrate the essential role of the surface Majorana flat band, we evaluate the Edelstein effect in a $s$-wave superconducting state by adopting 
$\Delta(\bm{k}) =\Delta_0$ instead of $\Delta(\bm{k}) =\Delta_0\sin k_x\sin k_y$. Keeping the other parameters in Fig.~\ref{fig:q_dept} we show the Edelstein coefficient $\gamma_{xy}$ in Fig.~\ref{fig:s-wave}.
Enhancement of the surface Edelstein effect is not observed in Fig.~\ref{fig:s-wave}, because the surface states do not appear in fullgap $s$-wave superconductors. 
Indeed, the Edelstein effect is nearly the same between the surface and bulk. This result supports the idea that the giant Edelstein effect in noncentrosymmetric $d$-wave superconductors is owing to the surface Majorana fermions. Note that the bulk Edelstein effect in a $d$-wave superconducting state (inset of Fig.~\ref{fig:q_dept}(d)) is an order of magnitude larger than that in a $s$-wave state (Fig.~\ref{fig:s-wave}). Therefore, the surface Edelstein effect in the former is approximately 1000 times larger than the superconducting Edelstein effect in a fully-gapped $s$-wave superconducting state.

\section{\label{sec:proposal}Conclusion and Proposal of experiments}
In this paper we investigated surface and bulk Edelstein effects in $d$-wave superconductors and demonstrated significant enhancement at surfaces due to the surface Majorana flat band. 
The surface Edelstein effect is typically 1000 times larger than the superconducting Edelstein effect in $s$-wave superconductors. This finding may pave a way for an efficient and energy-saving control of magnetic states by using topological surface states in superconductors. 

Finally, we would like to propose an experimental setup to realize and observe the giant Edelstein effect. Essential ingredients of the giant surface Edelstein effect are $d$-wave superconductivity, spacial inversion symmetry breaking, and spin-orbit coupling. 

    For materials, high-$T_{\rm c}$ cuprate superconductors are candidates.
    The $d$-wave superconductivity has been established \cite{scalapino1995case}, and space inversion asymmetry can be realized in heterostructures. For instance, fabrication of 
    ultra thin film by atomic-layer molecular beam epitaxy method has been reported~\cite{bollinger2011superconductor,PhysRevB.87.024509,PhysRevLett.107.027001}. Although a large spin-orbit coupling is not usually expected in 3$d$ electron systems, recent spin- and angle-resolved photoemission spectroscopy measurement for Bi2212 compounds observed a non-trivial spin texture with spin-momentum locking \cite{gotlieb2018revealing}, suggesting a sizable spin-orbit coupling. High transition temperature and large superconducting gap will be advantages for experimental studies for the superconducting Edelstein effect. 
    Another candidates are heavy fermion superlattices containing CeCoIn$_5$~\cite{mizukami2011extremely}. CeCoIn$_5$ is known to be a $d$-wave superconductor \cite{knebel2008quantum,pfleiderer2009superconducting} and space inversion symmetry breaking can be induced by an artificial control of superlattice structures~\cite{naritsuka2017emergent,shimozawa2016kondo}. 
    For the bulk noncentrosymmetric materials, CeRhSi$_3$~\cite{PhysRevLett.95.247004} and CeIrSi$_3$~\cite{sugitani2006pressure} are candidates of nodal spin-singlet superconductors. 
    A large spin-orbit coupling in heavy ions is expected to enhance the superconducting Edelstein effect in these heavy fermion superconductors.

    To observe a surface Edelstein effect enhanced by topological surface states it is important to choose an appropriate surface direction. 
    It is known that a topological winding number protecting the surface flat band is obtained as \cite{sato2011topology}
    \begin{equation}
    W(k_y) = \frac{1}{2}\sum_{\varepsilon(\bm{k})=0}\sum_{n} \mathrm{sgn}[\partial_{k_x}\varepsilon_n(\bm{k})]\cdot\mathrm{sgn}[\Delta_n(\bm{k})],
    \end{equation}
    with a band index $n$. Thus, the surface flat band appears when the gap function $\Delta_n(\bm{k})$ changes the sign in a direction normal to the surface. For high-$T_{\rm c}$ cuprate superconductors and heavy fermion superlattices $d_{x^2-y^2}$-wave superconductivity has been established. Then, the (110) surface is the best setting for a boundary condition. However, an enhanced Edelstein effect can be observed except for the (100) surface where the flat band disappears. 
    
    For a remark, the surface Edelstein effect may be suppressed by disorders because the uni-directional surface Majorana states can be delocalized by hybridizing with bulk states. 
    Thus, a clean surface is desirable for an experimental setup. Calculations of the Edelstein effect on disordered surfaces remain as a theoretical issue.
    
    To experimentally observe surface Edelstein effect, scanning SQUID is a possible experimental tool~\cite{kirtley1999scanning,wells2015analysis}. 
    For spintronics applications, magnetic domain switching in the superconductor/ferromagnet structure would be one of the goals of superconducting spintronics~\cite{linder2015superconducting}. In the setup shown in Fig.~\ref{fig:SC}, ferromagnetic domain switching can be caused by dissipationless supercurrent. Then, surface spin accumulation naturally occurs through the giant surface Edelstein effect. 

\acknowledgements
 The authors are grateful to A. Daido and S. Sumita for fruitful discussions and comments. This work was supported by KAKENHI (Grants No. JP15H05884, No. JP18H04225, No. JP18H05227, No. JP18H01178, and No. 20H05159) and by Core to Core program Oxide Superspin (OSS) international networking, from the Japan Society for the Promotion of Science (JSPS).
 






\nocite{*}
\bibliography{thesis}

\end{document}